\documentclass[aps, prd, twocolumn, nofootinbib]{revtex4-2}
\usepackage{amsmath}
\usepackage[colorlinks, linkcolor=red, anchorcolor=green, citecolor=blue]{hyperref}

\usepackage{graphicx}
\usepackage{epsfig}
\usepackage{bm}
\usepackage{color}
\usepackage{amsmath}
\usepackage{dcolumn}
\usepackage{enumitem}
\usepackage{orcidlink}

\usepackage{amssymb}
\begin{document}

\title{Anisotropic hydrodynamics with boost-non-invariant expansion}
\begin{abstract}
We establish the anisotropic hydrodynamics (aHydro) equations based on a boost-non-invariant longitudinally expanding system. Good consistency is found in the comparison between the aHydro results with those from the Boltzmann equation under relaxation time approximation. 
We also obtain corresponding attractor solution and observe the time delay of convergence when increasing the absolute value of the rapidity. We finally show that the rapidity dependence is solved by introducing the redefined attractor variables to compensate the time delay effect.
\end{abstract}
\author{Shile Chen} \email{csl2023@tsinghua.edu.cn}
\affiliation{Department of Physics, Tsinghua University, Beijing 100084, China}
\author{Shuzhe Shi} \email{shuzhe-shi@tsinghua.edu.cn}
\affiliation{Department of Physics, Tsinghua University, Beijing 100084, China}
\maketitle
\section{Introduction}

Relativistic heavy-ion collisions are pursued at the Relativistic Heavy-Ion Collider (RHIC) and at the Large Hadron Collider (LHC) to explore the nuclear matter properties at high energy density.
A color-deconfined phase of matter -- the Quark-Gluon Plasma (QGP) --- is found and studied systematically. One of the breakthroughs in theoretical relativistic heavy-ion physics has been the great success of numerical hydrodynamic simulations in understanding and predicting measurements of the collective behavior of the final state hadrons (see e.g. Refs.~\cite{Shen:2014vra, Schenke:2010rr, Karpenko:2013wva, vanderSchee:2013pia, Pang:2018zzo, Du:2019obx}), which shows the applicability of hydrodynamics after a surprisingly short time after the initial collision. The hydrodynamics, based on the gradient expansion of macroscopic limit requires the system near local equilibrium, which indicates the thermalization in the microscopic distribution function. Since the initial condition of the relativistic heavy-ion collisions is far from equilibrium, the reason of unexpected rapid hydrodynamization in relativistic heavy-ion collisions remains unsolved, and many efforts have been made~\cite{Heller:2011ju, Heller:2013fn,  Heller:2015dha, Kurkela:2015qoa, Blaizot:2017lht, Romatschke:2017vte, Spalinski:2017mel, Romatschke:2017acs, Behtash:2017wqg, Blaizot:2017ucy, Romatschke:2017ejr, Kurkela:2018wud, Mazeliauskas:2018yef, Behtash:2019txb, Heinz:2019dbd, Blaizot:2019scw, Blaizot:2020gql, Blaizot:2021cdv, Martinez:2010sc, Ryblewski:2010ch, Martinez:2012tu, Ryblewski:2012rr, Strickland:2017kux, Strickland:2018ayk, Brewer:2019oha, Brewer:2022ifw}.
Among them, the anisotropic hydrodynamics theory (aHydro)~\cite{Martinez:2010sc, Ryblewski:2010ch, Martinez:2012tu, Ryblewski:2012rr} has been constructed to take into account the longitudinal versus transverse anisotropy in the distribution function. The aHydro is useful in studying the isotropization --- an important cricteria of thermalization --- of a far-from-equilibrium system.

Meanwhile, the attractor solutions have been found for M\"uller--Isreal--Stewart hydrodynamics~\cite{Heller:2011ju, Heller:2013fn, Heller:2015dha} and for aHydro~\cite{Strickland:2017kux, Strickland:2018ayk}. 
In such solutions assuming boost-invarance in the longitudinal direction and translation-invariance in the transverse plane, initial-state-independent common behavior was observed with respect to the time evolution scaled by relaxation time. The existence of attractors implies universal effective theory before equilibration, and it might help in understanding the rapid applicability of hydrodynamic theory.
Note that realistic heavy-ion collisions have non-trivial rapidity distribution. Testing the applicability of hydrodynamics in the longitudinal direction provides an extra information in studying early evolution of such a far-from-equilibrium system.
In this work, we relieve the boost-invariance assumption by going to the comoving frame of an analytical solution with non trivial rapidity dependence~\cite{Shi:2022iyb}.
We construct the aHydro equations in such a frame, compared to the result of Boltzmann equation, and also find the attractor.

This paper is organized as follows. In Sec.~\ref{sec:ahydro}, we construct the boost-non-invariant aHydro equations and then present the numerical solutions in comparison with those of Boltzmann equations. Then we study the attractor in Sec.~\ref{sec:attractor} and finally summarize in Sec.~\ref{sec:summary}.

\section{Construction of aHydro equations}\label{sec:ahydro}
\subsection{Comoving frame of asymmetric hydro solution}
For the sake of convenience, we construct the anisotropic hydro equations in the comoving frame of the analytical solution found in Ref.~\cite{Shi:2022iyb}. Their transformation with Milne coordinate is 
\begin{align}
\begin{split}
\hat{\tau} =\;& 
    \frac{2a\,\tau_i}{1+a^2}\Big(
    \big(\frac{t_0 + a\,\tau e^{\eta}}{\tau_i}\big)^{\frac{1}{a}}
    \big(\frac{{t}_0}{\tau_i} + \frac{\tau e^{-\eta}}{a\,\tau_i}\big)^a 
    \Big)^{\frac{1+a^2}{4a}} \,,\\
\hat{\eta} =\;& 
    \frac{1+a^2}{4a}\ln\Big(
    \big(\frac{t_0 + a\,\tau e^{\eta}}{\tau_i}\big)^{\frac{1}{a}}
    \big/
    \big(\frac{{t}_0}{\tau_i} + \frac{\tau e^{-\eta}}{a\,\tau_i}\big)^a 
    \Big) \,,\\
\hat{x}^x =\;& x,\qquad\qquad \hat{x}^y=y.
\end{split}
\label{eq:coordinate}
\end{align}
where $\tau$ and $\eta$ are the proper-time and rapidity in Milne coordinates, respectively. $a\in (1/\sqrt{2},\sqrt{2})$ is a dimensionless parameter characterizing the asymmetry between forward and backward rapidity range and the system would return symmetric when taking $a=1$. We keep the notation of general $a$ until at a point that we will show one should set $a=1$ to obtain the aHydro equations. The non-negative time constant $t_0$ serves as a translation of the Minkowski time, and it corresponds to the time needed for the colliding nuclear pancakes to pass through each other in relativistic heavy-ion collisions~\cite{Shi:2022iyb}.
$\tau_i$ is a free parameter of the time unit, and we set it as the initial time of the dynamical evolution. In what follows, we employ the hat notation ($\hat{}$) to denote variables in the comoving frame, and un-dressed symbols as those in the Milne frame.

In the co-moving frame, the metric reads
\begin{align}
\hat{g}^{\mu\nu} =\;& \mathrm{diag}(\hat{g}^{\tau\tau}, \hat{g}^{\eta\eta}, -1, -1)\,,\\
\hat{g}_{\mu\nu} =\;& \mathrm{diag}(\frac{1}{\hat{g}^{\tau\tau}}, \frac{1}{\hat{g}^{\eta\eta}}, -1, -1)\,,\\
\hat{g}^{\tau\tau} =\;&
    e^{2\frac{1-a^2}{1+a^2} \hat{\eta}}
    \,,\qquad
\hat{g}^{\eta\eta} =
     - \frac{1}{\hat{\tau}^2}
    e^{2\frac{1-a^2}{1+a^2} \hat{\eta}}\,,
\end{align}
and the nonvanishing elements of the corresponding Christopher symbols are
\begin{align}
\begin{split}
&   
    \hat{\Gamma}^{\tau}_{\;\,\eta\eta} = \hat \tau\,,\quad
    \hat{\Gamma}^{\eta}_{\;\,\eta\tau} = \hat{\Gamma}^{\eta}_{\;\,\tau\eta} = \frac{1}{\hat \tau}\,,
\\
&    \hat{\Gamma}^{\tau}_{\;\,\hat \tau\eta} = \hat{\Gamma}^{\tau}_{\;\,\eta\tau} 
=    \hat{\Gamma}^{\eta}_{\;\,\eta\eta} 
= \hat \tau^2\,\hat{\Gamma}^{\eta}_{\;\,\tau\tau}
=    \frac{a^2-1}{a^2+1}\,.
\end{split}
\end{align}

The spatial components of fluid velocity in solution~\cite{Shi:2022iyb} are vanishing, and the solution is given by
\begin{align}
\frac{\hat{T}_\mathrm{ideal}}{T_0} =\;& \sqrt{\hat{g}^{\tau\tau}} \Big(\frac{\tau_i}{\hat{\tau}}\Big)^{\frac{1}{3}}\,,\\
\hat{u}^{\tau} =\;& \sqrt{\hat{g}^{\tau\tau}}\,,
\qquad
\hat{u}^{\eta} = \hat{u}^x = \hat{u}^y = 0\,.
\end{align}
and finally, the onshell condition of a massless particle reads
\begin{align}
\begin{split}
%\hat{p}_{\tau} =\;& e^{-\frac{1-a^2}{1+a^2} \hat{\eta}} \sqrt{-\hat{g}^{\eta\eta}\hat p_\eta^2+ p_x^2+p_y^2}\,,\\
\hat u\cdot \hat p =\;& \sqrt{ -\hat{g}^{\eta\eta}\hat{p}_{\eta}^2+ \hat{p}_x^2+\hat{p}_y^2}\,.
\end{split}
\end{align}

\subsection{Equations of motion for distribution function and moments}
Taking the comoving frame, the Boltzmann equation for on-shell distribution function reads, (see e.g., ~\cite{Hohenegger:2008zk, Lee:2012sd} and ~\cite{Chen:2023vrk})
\begin{align}
\begin{split}
&
    \Big(\hat{p}^{\tau}\frac{\partial}{\partial{\hat{\tau}}} 
    + \hat{p}^{\eta}\frac{\partial}{\partial \hat{\eta}} 
    + \frac{a^2-1}{a^2+1} (\hat p_x^2+\hat p_y^2) \frac{\partial}{\partial{\hat{p}_\eta}} \Big) f(\hat x,\hat p) = \mathcal{C}[f]\,,
\\
&\mathcal{C}[f]\equiv 
     -\frac{\hat{p}_{\tau} \hat{u}^{\tau} \hat{T}(\hat{\tau})}{5\bar \eta}\Big(f(\hat{x},\hat{p})-f_\mathrm{eq}(\hat{x},\hat{p})\Big)\,,
\end{split}
\label{eq:boltzmann}
\end{align}
where $\frac{5\bar\eta}{\hat{T}(\hat{x})}$ is the local relaxation time and equilibrium distribution is $f_\mathrm{eq} = e^{-\hat{p}\cdot \hat{u}/\hat{T}}$.
We further define the moments of the collision kernel as
\begin{align}
    \mathcal{C}_{r}^{\mu_1\cdots\mu_n} \equiv 
    \int d\hat{P} (\hat{p}\cdot \hat{u})^r \hat{p}^{\mu_1} \cdots \hat{p}^{\mu_n} \mathcal{C}[f]\,,
\end{align}
where $d \hat P = \frac{\hat{g}^{\tau\tau}}{\hat{\tau}} \frac{d\hat{p}_{\eta} d\hat{p}_x d\hat{p}_y}{(2\pi)^3 \hat{p}^{\tau}}$ denotes the momentum integration measurement.
The continuity equations of the charge currents, energy-momentum tensor, and second order tensor,
\begin{align}
\begin{split}
    \hat{J}^\mu =\;& \int d\hat{P}\, \hat{p}^\mu f\,,\\
    \hat{T}^{\mu\nu} =\;& \int d\hat{P}\, \hat{p}^\mu \hat{p}^\nu f\,,\\
    \hat{I}^{\mu\nu\lambda} =\;& \int d\hat{P}\, \hat{p}^\mu \hat{p}^\nu \hat{p}^\lambda f\,,
\end{split}
\end{align}
would follow
\begin{align}
\begin{split}
    \hat{\mathcal{D}}_\mu \hat{J}^\mu =\;& \mathcal{C}_0\,,\\
    \hat{\mathcal{D}}_\mu \hat{T}^{\mu\nu} =\;& \mathcal{C}_0^{\nu}\,,\\
    \hat{\mathcal{D}}_\mu \hat{I}^{\mu\nu\lambda} =\;& \mathcal{C}_0^{\nu\lambda}\,.
\end{split}\label{eq:evolution_momentum}
\end{align}
Here $\hat{\mathcal{D}}_\mu$ is the covariant derivative. For instance, 
$\hat{\mathcal{D}}_\mu \hat{V}^\nu = \hat{\partial}_\mu \hat{V}^\nu + \hat{\Gamma}^\nu_{\ \mu\sigma} \hat{V}^\sigma$ and 
$\hat{\mathcal{D}}_\mu \hat{T}^{\mu\nu} = \hat{\partial}_\mu \hat{T}^{\mu\nu} + \hat{\Gamma}^{\mu}_{\;\,\rho\mu}\hat{T}^{\rho\nu} + \hat{\Gamma}^{\nu}_{\;\,\rho\mu}\hat{T}^{\rho\mu}$.

\subsection{aHydro Equations}
Noting that particles created in the initial process of heavy-ion collisions is highly anisotropic when one compares the longitudinal and transverse distributions, anisotropic hydro (aHydro) framework takes the Romatschke--Strickland (RS)~\cite{Strickland:2017kux,Strickland:2018ayk} form to study thermalization, especially the isotropization,
\begin{align}
    f(\hat{x}, \hat{p}) = f_\mathrm{RS}(\hat{x}, \hat{p}) = e^{-\frac{\sqrt{(\frac{\hat p_\eta}{\hat{\tau}})^2(\xi(\hat{x})+1)+ \hat{g}_{\tau\tau}(\hat{p}_x^2 + \hat{p}_y^2)}}{\Lambda(\hat{x})}}\,,
    \label{eq:RSdistribution}
\end{align}
where $\xi(\hat{x})$ characterize the degree of anisotropy. $\Lambda(\hat{x})$ is the scale of the energy and has the meaning of $\frac{\hat T}{\sqrt{\hat{g}^{\tau\tau}}}$ when the distribution is isotropic ($\xi=0$). 

Taking the RS form~\eqref{eq:RSdistribution}, the stress tensor reads
\begin{align}
\begin{split}
\hat{T}^{\tau\tau}(\xi,\Lambda) 
=\;&
    \frac{1}{2}\Big(\frac{1}{1+\xi}+\frac{\arctan\sqrt{\xi}}{\sqrt{\xi}}\Big)\times(\hat{g}^{\tau\tau})^3 
    \frac{3\Lambda^4(\hat{x})}{\pi^2}
\\\equiv\;&
    \mathcal R(\xi) \times \hat{T}^{\tau\tau}_\mathrm{eq}(\Lambda)\,,
\end{split}\label{eq:T00}\\
\begin{split}
\hat{T}^{\eta\eta}(\xi,\Lambda) 
=\;&
    \frac{3}{2}\Big(\frac{\arctan\sqrt{\xi}}{(\sqrt{\xi})^3}-\frac{1}{\xi(\xi+1)}\Big)\times\frac{(\hat{g}^{\tau\tau})^3}{(\hat{\tau})^2} 
    \frac{\Lambda^4(\hat{x})}{\pi^2}
\\\equiv\;&
    \mathcal R_\mathrm{L}(\xi) \times \hat{T}^{\eta\eta}_\mathrm{eq}(\Lambda)\,,
\end{split}\label{eq:T11}\\
\begin{split}
\hat{T}^{xx}(\xi,\Lambda) 
=\;&
    \frac{3}{4}\Big(\frac{(\xi-1)\arctan\sqrt{\xi}}{(\sqrt{\xi})^3}+\frac{1}{\xi}\Big)
    \times (\hat{g}^{\tau\tau})^2 
    \frac{\Lambda^4(\hat{x})}{\pi^2}
\\\equiv\;&
    \mathcal R_\mathrm{T}(\xi) \times \hat{T}^{xx}_\mathrm{eq}(\Lambda)\,,
\end{split}\label{eq:Txx}
\end{align}
and other components vanish except for $\hat{T}^{yy}(\xi,\Lambda) = \hat{T}^{xx}(\xi,\Lambda)$. Regarding the third order tensors, we focus on the temporal-spatial-spatial components in order to study pressure anisotropy~\cite{Alqahtani:2017mhy}. They read
\begin{align}
\begin{split}
\hat{I}^{\tau\eta\eta}(\xi,\Lambda)
=\;&
    \frac{1}{(1+\xi)^{\frac{3}{2}}}
    \times  
    \frac{(\hat{g}^{\tau\tau})^4}{\hat{\tau}^2} \frac{12\Lambda^5}{\pi^2}
\\\equiv\;&
    \mathcal S_\mathrm{L}(\xi) \times I^{\tau\eta\eta}_\mathrm{eq}(\Lambda)\,,
\end{split}\\
\begin{split}
\hat{I}^{\tau xx}(\xi,\Lambda)
=\;&
    \frac{1}{\sqrt{1+\xi}}
    \times (\hat{g}^{\tau\tau})^2 \frac{4\Lambda^5}{\pi^2}
\\\equiv\;&
    \mathcal S_\mathrm{T}(\xi) \times \frac{\hat{\tau}^2}{(\hat{g}^{\tau\tau})^2} I^{\tau\eta\eta}_\mathrm{eq}(\Lambda)\,.
\end{split}
\end{align}
The equilibrium forms of stress tensor and third order tensor are obtained by simply taking the isotropic limit that $\xi=0$, and the anisotropy correction is carried by the factors $
\mathcal R(\xi) = \frac{1}{2}\big(\frac{1}{1+\xi}+\frac{\arctan\sqrt{\xi}}{\sqrt{\xi}}\big)$,
$\mathcal R_\mathrm{L}(\xi) = \frac{(\xi+1)\mathcal R(\xi)-1}{(\xi^2+\xi)/3}$, and $\mathcal R_\mathrm{T}(\xi) = \frac{(\xi^2-1)\mathcal R(\xi)+1}{2(\xi^2+\xi)/3}$ for stress tensor, as well as
 $\mathcal S_\mathrm{L}(\xi) = (1+\xi)^{-\frac{3}{2}}$ and $\mathcal S_\mathrm{T}(\xi) = (1+\xi)^{-\frac{1}{2}}$ for the third order tensor. 

For the moments of collision kernels, we find
\begin{align}
% \begin{split}
% \mathcal{\hat{C}}_0^{\tau\tau} =\;& 
%     \frac{\mathcal{R}^{\frac{1}{4}}\Lambda}{5\,\bar\eta} \big(\mathcal R^{\frac{5}{4}}(\xi)- \mathcal S(\xi)\big) I_\mathrm{eq}^{\tau\tau\tau}(\Lambda)\,,
% \end{split}\\%~\nonumber\\
\begin{split}
\mathcal{\hat{C}}_0^{\eta\eta} =\;& 
    \frac{\mathcal{R}^{\frac{1}{4}}\Lambda}{15\,\bar\eta} \big(\mathcal R^{\frac{5}{4}}(\xi)- \mathcal S_\mathrm{L}(\xi)\big) I_\mathrm{eq}^{\tau\eta\eta}(\Lambda)\,,
\end{split}\\%~\nonumber\\
\begin{split}
\mathcal{\hat{C}}_0^{xx} =\;& 
    \frac{\mathcal{R}^{\frac{1}{4}}\Lambda}{15\,\bar\eta} \big(\mathcal R^{\frac{5}{4}}(\xi)- \mathcal S_\mathrm{T}(\xi)\big) \frac{\hat{\tau}^2}{(\hat{g}^{\tau\tau})^2} I^{\tau\eta\eta}_\mathrm{eq}(\Lambda)\,,
\end{split}
\end{align}
for the third order tensors, and $\mathcal{\hat{C}}_0^{\mu} = 0$ to assure energy-momentum conservation. The latter can be fulfilled by imposing the Landau matching condition that $T^{\tau\tau}(\xi,\Lambda)= T^{\tau\tau}_\mathrm{eq}(\frac{\hat T}{\sqrt{\hat{g}^{\tau\tau}}})$, which means
\begin{align}
\hat T(\hat{x}) = \sqrt{\hat{g}^{\tau\tau}}\, \mathcal R^{\frac{1}{4}}(\xi(\hat{x})) \, \Lambda(\hat{x})\,.
\label{eq:ahydro:temperature}
\end{align}
Meanwhile, the pressure anisotropy, defined as the ratio between longitudinal and transverse pressures, is of interest.
It can be represented by $\xi$ according to
\begin{align}
    \frac{P_\mathrm{L}}{P_\mathrm{T}} 
\equiv \frac{-\hat{T}^{\eta\eta}(\xi,\Lambda)/\hat{g}^{\eta\eta}}{-\hat{T}^{xx}(\xi,\Lambda)/\hat{g}^{xx}} = 
    \frac{\mathcal R_\mathrm{L}(\xi)}{\mathcal R_\mathrm{T}(\xi)}\,.
\label{eq:ahydro:pressure}
\end{align}

If keeping all orders in the series, the extension of equation set~\eqref{eq:evolution_momentum} would be equivalent to the Boltzmann equation~\eqref{eq:boltzmann}. Nevertheless, since the RS form~\eqref{eq:RSdistribution} does not follow~\eqref{eq:boltzmann} at all time, not all equations in~\eqref{eq:evolution_momentum} can be satisfied. Therefore, as an effective theory, the aHydro framework only takes into account the equations related to the quantities of interested --- temperature and pressure anisotropy~\cite{Alqahtani:2017mhy}. 
Relevant equations are the energy conservation $\hat{\mathcal{D}}_\mu \hat{T}^{\mu\tau} = 0$, and the longitudinal-transverse difference $\hat{\mathcal{D}}_\mu(\hat{I}^{\mu\eta\eta}-\hat{I}^{\mu xx})$, and their explicit expression read 
\begin{align}
    \hat{\partial}_\tau \hat{T}^{\tau\tau}
    +\frac{\hat{T}^{\tau\tau}}{\hat{\tau}}
    +\hat{\tau} \hat{T}^{\eta\eta} =\;& 0
    \,,\\
\hat{\partial}_\tau \big(\hat{I}^{\tau\eta\eta} - \hat{I}^{\tau xx} \big)
    + \frac{5}{\hat{\tau}} \hat{I}^{\tau\eta\eta}
    - \frac{1}{\hat{\tau}} \hat{I}^{\tau xx} =\;& \mathcal{\hat{C}}_0^{\eta\eta}-\mathcal{\hat{C}}_0^{xx}\,,
\end{align}
and they respectively lead to
\begin{align}
&\frac{\mathcal R'(\xi)}{\mathcal R(\xi)} \hat{\partial}_\tau\xi+\frac{4}{\Lambda}\hat{\partial}_\tau\Lambda = \frac{1}{\hat{\tau}}\Big(\frac{1}{\xi(1+\xi)\mathcal R(\xi)}-\frac{1}{\xi}-1\Big)\,, \label{eq:eom:1st}\\
&-\frac{1}{1+\xi} \hat{\partial}_\tau\xi+\frac{2}{\hat{\tau}}
=  \frac{\xi\sqrt{1+\xi}}{5\,\bar\eta}\mathcal R^{\frac{3}{2}}\,\Lambda\,,
\label{eq:eom:2nd}
\end{align}
with $\mathcal R'(\xi) = \frac{\mathrm{d}\mathcal R(\xi)}{\mathrm{d}\xi}$. So far, such equations reproduce those in Ref.~\cite{Alqahtani:2017mhy}, although they have different meanings. With trivial rapidity dependence, the equation of motion for the longitudinal components is automatically satisfied in Ref.~\cite{Alqahtani:2017mhy}. Nevertheless, it is not the case in the present work.
The longitudinal momentum conservation ($\hat{\mathcal{D}}_\mu \hat{T}^{\mu\eta} = 0$) gives
\begin{align}
0=\;&
    \hat{\partial}_\eta \hat{T}^{\eta\eta}
    +3 \frac{a^2-1}{a^2+1}\hat{T}^{\eta\eta} 
    +\frac{1}{\hat{\tau}^2} \frac{a^2-1}{a^2+1} \hat{T}^{\tau\tau}
    \,,
\end{align}
which can be simplified as
\begin{align}
    \frac{\mathcal{R}_\mathrm{L}'(\xi)}{\mathcal{R}_\mathrm{L}(\xi)}\hat{\partial}_\eta\xi+\frac{4}{\Lambda}\hat{\partial}_\eta\Lambda+ 3 \frac{1-a^2}{1+a^2}
    \Big(1-\frac{\mathcal{R}(\xi)}{\mathcal{R}_\mathrm{L}(\xi)}\Big)=0\,.
    \label{eq:aHydro_eta}
\end{align}
Apparently, the rapidity dependence of $\xi$ and $\Lambda$ is non-trivial unless for symmetric system ($a=1$) or when the system approaches the isotropic limit that $\xi\to0$ so that both $\mathcal{R}(\xi)$ and $\mathcal{R}_\mathrm{L}(\xi)$ approach unity. For general $a$, Eq.~\eqref{eq:aHydro_eta} would lead to a relation that contradicts to (\ref{eq:eom:1st}, \ref{eq:eom:2nd}), i.e.,  $\hat{\partial}_\tau \hat{\partial}_\eta \Lambda = \hat{\partial}_\eta \hat{\partial}_\tau \Lambda$ and $\hat{\partial}_\tau \hat{\partial}_\eta \xi = \hat{\partial}_\eta \hat{\partial}_\tau \xi$ cannot be simultaneously satisfied. Thus, we conclude that aHydro equations taking the RS form~\eqref{eq:RSdistribution} only apply to the symmetric case ($a=1$). In such case, $\Lambda$ and $\xi$ only depend on $\hat{\tau}$ but not $\hat{\eta}$.

\section{Comparison with the Boltzmann equation}
To check the validity of aHydro, we numerically solve Eqs.~(\ref{eq:eom:1st}, \ref{eq:eom:2nd}) and compare the observables with those from Boltzmann equations.
First, we simplify the equations as
\begin{align}
\hat{\partial}_\tau\Lambda =\;&
\frac{\xi(1+\xi)^{\frac{3}{2}}\mathcal R^{\frac{1}{2}}(\xi) \mathcal R'(\xi)\, \Lambda^2}{20\bar\eta}
\,,\\
\hat{\partial}_\tau\xi =\;& \frac{2(1+\xi)}{\hat{\tau}} -\frac{\xi(1+\xi)^{\frac{3}{2}}\mathcal R^{\frac{3}{2}}(\xi) \Lambda}{5\bar\eta}\,,
\label{eq:aHydro_eq1}
\end{align}
with
\begin{align}
& \mathcal R(\xi) = \frac{1}{2}\Big(\frac{1}{1+\xi}+\frac{\arctan\sqrt{\xi}}{\sqrt{\xi}}\Big)\,.
\end{align}
They further yield that
\begin{align}
    \hat{\tau} \hat{\partial}_\tau \ln (\mathcal{R} \Lambda^4) = \frac{2\mathcal{R}'}{\mathcal{R}}(1+\xi)\,.
\label{eq:aHydro_eq2}
\end{align}
We solve the numerical equations with initial conditions $\xi(\hat{\tau}=\tau_i) = \xi_0$ and $\Lambda(\hat{\tau}=\tau_i) = \Lambda_0$.
With $\Lambda$ and $\xi$, one may further compute observables of interest, i.e., temperature and pressure anisotropy according to Eq.~\eqref{eq:ahydro:temperature} and~\eqref{eq:ahydro:pressure}, respectively.

Before solving the equations numerically, we first analyze the long-time asymptotic behaviors. When $\hat{\tau} \to \infty$, the system approaches the isotropic limit, and we expand in series of $\xi$ around zero and keep up to second order. The equations become
\begin{align}
\hat{\partial}_\tau \ln (\mathcal{R}^{\frac{1}{4}} \Lambda) =& 
    -\frac{1}{3\,\hat{\tau}} 
    + \frac{2}{45}\frac{\xi}{\hat{\tau}} 
    - \frac{26}{945}\frac{\xi^2}{\hat{\tau}} 
    + \mathcal{O}(\xi^3),\\
\hat{\partial}_\tau\xi =& 
\frac{2(1+\xi)}{\hat{\tau}} - (\mathcal{R}^{\frac{1}{4}} \Lambda) \Big(\frac{\xi}{5} + \frac{2\xi^2}{15} \Big) \,,
\end{align}
and we find the long time behavior as
\begin{align}
\mathcal{R}^{\frac{1}{4}}\Lambda = \frac{c_0}{\hat{\tau}^{\frac{1}{3}}} - \frac{2\bar\eta}{3\hat{\tau}}\,,\quad
\xi = \frac{10\bar\eta}{c_0} \hat{\tau}^{-\frac{2}{3}}
+ \frac{200\bar\eta^2}{3c_0^2} \hat{\tau}^{-\frac{4}{3}}\,.
\label{eq:asymptotic}
\end{align}
The leading terms correspond to the Bjorken flow with temperature scaled by a constant greater than unity because of entropy production. Such rescaling is encoded in the constant $c_0$ that depends on the initial condition and can only be determined by fitting the numerical solution.

\begin{figure}[!hbpt]
    \centering
    \includegraphics[width=0.4\textwidth]{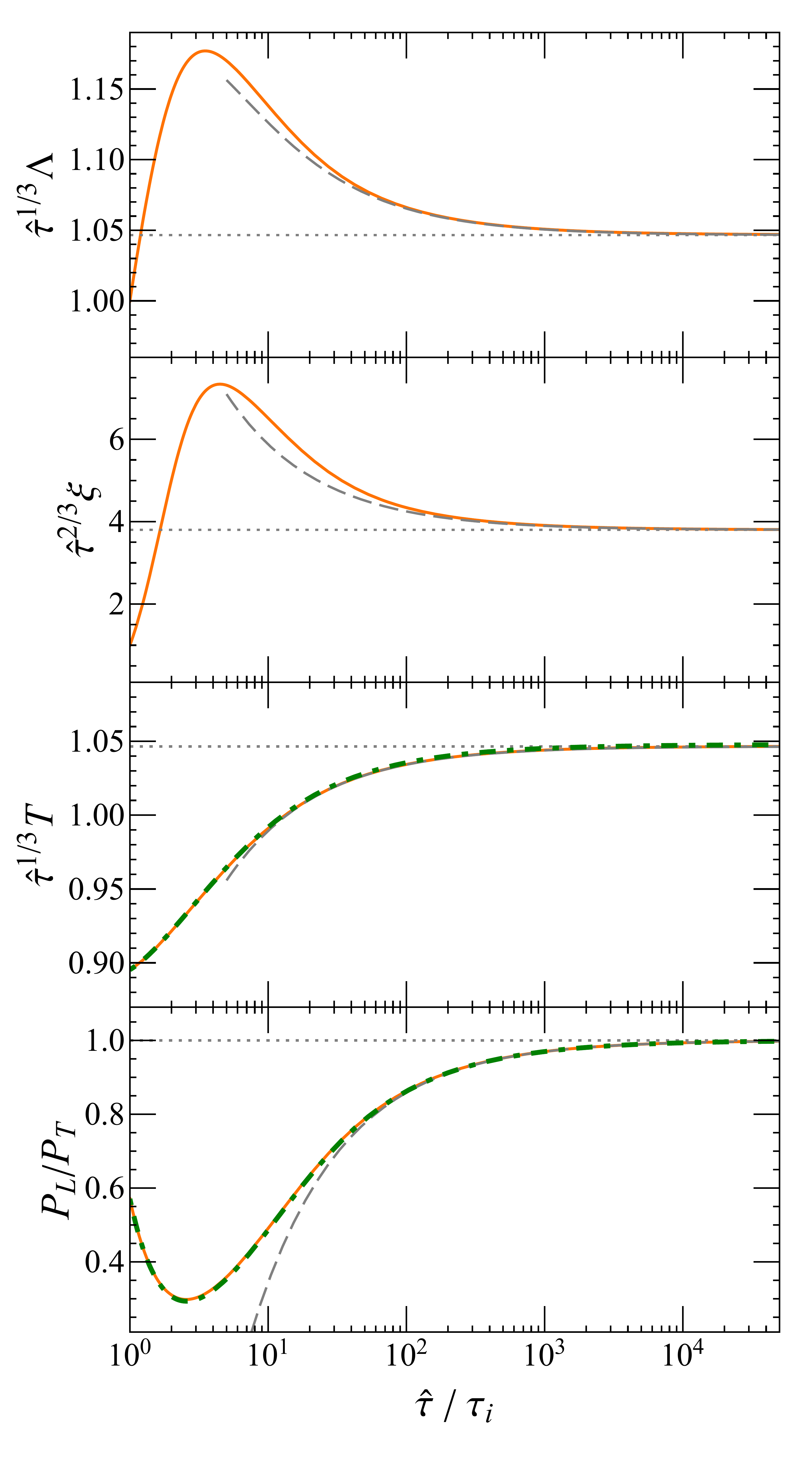}
    \caption{(From top to bottom) Hat-proper time evolution of $\Lambda$, $\xi$, temperature, and longitudinal-to-transverse-pressure-ratio. Orange curves are results from aHydro, gray dotted and dashed curves are respectively asymptotic results up to the leading order and the next to leading order, green dash-dotted curves are from formal solution of the RTA Boltzmann equation.
    \label{fig1}}
\end{figure}
%\subsection{Results from RTA Boltzmann equation}
To check the applicability of aHydro framework, it would be important to compare the observables of interest from aHydro with those computed from Boltzmann equation~\eqref{eq:boltzmann}.
Eq.~\eqref{eq:boltzmann} has no analytical solution in general. However, in the special case that $a=1$, one may find a formal solution~\cite{Baym:1984np, Florkowski:2013lya, Florkowski:2013lza},
\begin{align}
\label{eq:boltzmann_formal_solution}
\begin{split}
    f(\hat{\tau}, \hat{p}_{\tau}) 
=\;&  
    D(\hat{\tau}, \tau_i)f_0(\tau_i, \hat p_{i,\tau})\\
+\;&
    \frac{1}{5\bar{\eta}} \int_{\tau_i}^{\hat{\tau}} d{\hat x}' D(\hat{\tau}, {\hat x}' ) \hat{T}({\hat x}' )f_\mathrm{eq}({\hat x}', \hat{p}_{\tau}')\,,
\end{split}
\end{align}
with
\begin{align}
    D(\hat{\tau}, \hat{\tau}' ) \equiv \exp\Big(-\frac{1}{5\bar\eta} \int_{\hat{\tau}'}^{\hat{\tau}} T(t) \,dt\Big)\,.
\end{align}
For a fair comparison, the initial distribution in the Boltzmann equation is taken to be identical to that of the aHydro, i.e., $f_0(\tau_i,\hat p_{i,\tau}) = e^{-\frac{\sqrt{(\frac{\hat{p}_\eta}{\tau_i})^2(\xi_0+1)+ \hat{p}_x^2 + \hat{p}_y^2}}{\Lambda_0}}$. Then the formal solution of the RTA Boltzmann equation~\eqref{eq:boltzmann_formal_solution} gives that~\cite{Chen:2023vrk}
\begin{align}
\begin{split}
 \;&    \hat{T}^{\mu\nu}(\hat{\tau})
=
    D(\hat{\tau}, \tau_i) 
    \hat{T}_0^{\mu\nu}(\hat{\tau})
\\+\;&
    \frac{1}{5\bar{\eta}} \int_{\tau_i}^{\hat{\tau}} d{\hat x}' D(\hat{\tau}, {\hat x}' ) \hat{T}({\hat x}') 
    \hat{T}^{\mu\nu}\big(({\hat{\tau}}/{\hat x'})^2-1, \hat{T}(\hat x')\big),
\end{split}
\label{eq:Tmunu}
\end{align}
and the initially induced part is given by
\begin{align}
    \hat{T}_0^{\mu\nu}(\hat{\tau}) =  \hat{T}^{\mu\nu}\Big((1+\xi_0)\big(\frac{\hat{\tau}}{\tau_i}\big)^2 - 1, \Lambda_0\Big)\,,
\end{align}
with $\hat{T}^{\mu\nu}(\xi,\Lambda)$ given in Eqs.~(\ref{eq:T00}--\ref{eq:Txx}). The effective temperature is given by the self-consistent equation
\begin{align}
\begin{split}
    \hat{T}^4(\hat{\tau}) =\;& 
    D(\hat{\tau}, \tau_i) \frac{\mathcal{R}\Big((1+\xi_0)\big(\frac{\hat{\tau}}{\tau_i}\big)^2 - 1\Big)}{\mathcal{R}(\xi_0)}T_0^4 
\\&
     +\int_{\tau_i}^{\hat{\tau}} 
     \frac{d{\hat x}'}{5\bar{\eta}}
     D(\hat{\tau}, {\hat x}') \hat{T}^5({\hat x}') 
    \mathcal{R}\big(({\hat{\tau}}/{\hat x'})^2-1\big)\,,
\end{split}\label{eq:RTA:temperature}
\end{align}
and the longitudinal and transverse pressures read
\begin{align}
\begin{split}
     \hat{P}_{j}(\hat{\tau})
=\;&
    D(\hat{\tau}, \tau_i)
    \frac{\mathcal{R}^{j} ((1+\xi_0)\big(\frac{\hat{\tau}}{\tau_i}\big)^2 - 1)}{\mathcal{R}(\xi_0)} \frac{T_0^4}{\pi^2}
\\+\;&
     \int_{\tau_i}^{\hat{\tau}} \frac{d{\hat x}'}{5\bar{\eta}} D(\hat{\tau}, {\hat x}' ) \frac{\hat{T}^5({\hat x}')}{\pi^2}
    \mathcal{R}_{j}(({\hat{\tau}}/{\hat x'})^2-1),
\end{split}\label{eq:RTA:pressure}
\end{align}
with $j \in \{ \mathrm{L}, \mathrm{T} \}$.

\begin{figure*}
    \centering
    \includegraphics[width=0.45\textwidth]
    {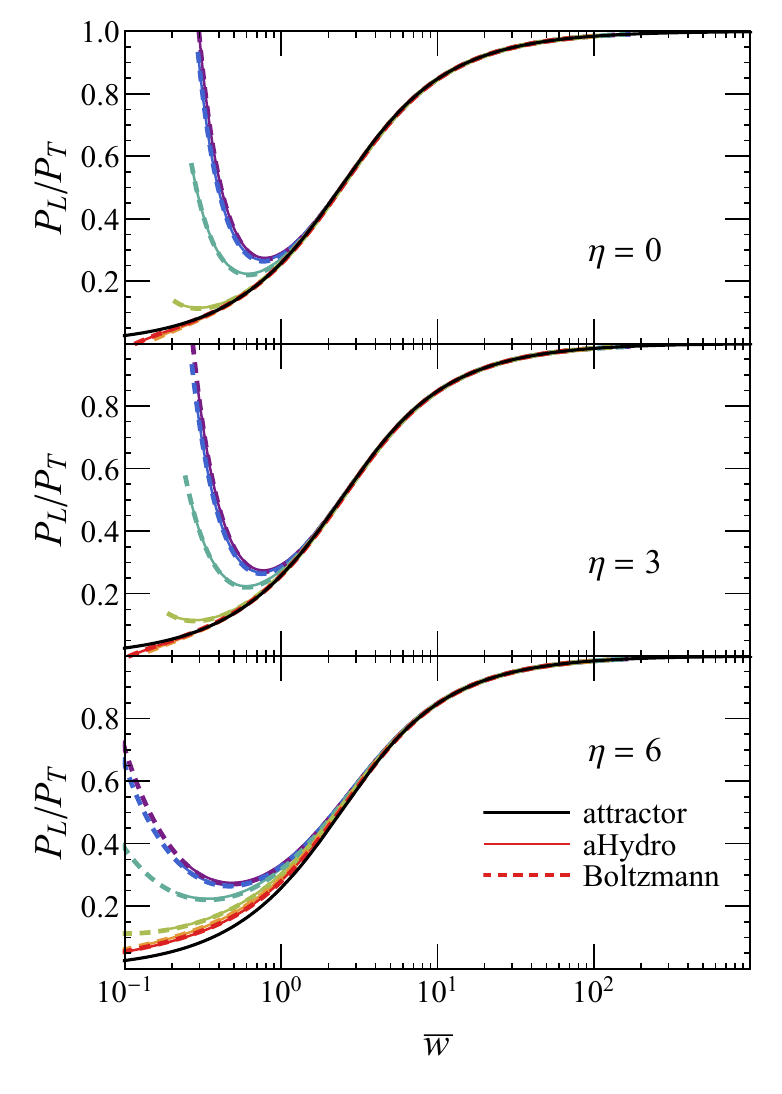}\quad
    \includegraphics[width=0.45\textwidth]
    {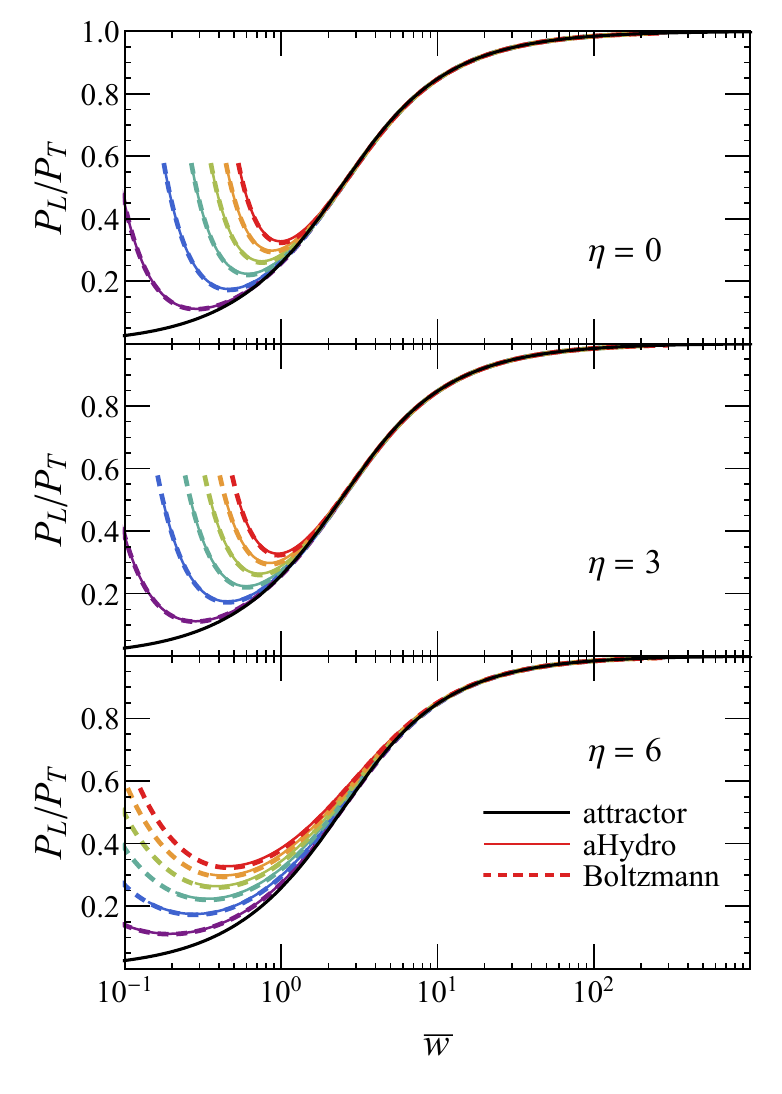}
    \caption{AHydro solution (solid rainbow lines) in comparison with Boltzmann solution (dashed rainbow lines) of the pressure anisotropy ($P_\mathrm{L}/P_\mathrm{T}$). From top to bottom are for $\eta = 0$, $3$ and $6$, respectively.
    In the left panel, we fix that $\Lambda_0 = 0.6$, and from purple to red $\xi_0$ varying $\{10^{-2}, 10^{-1}, 10^{0}, 10^{1}, 10^{2}, 10^{3}\}$; 
    in the right panel, we fixed $\xi_0 = 1.0$ and from purple to red respectively correspond to $\Lambda_0 = \{0.2, 0.4, 0.6, 0.8, 1.0, 1.2\}$.
    In all panels, the black thick solid lines represent the attractor solution.}
    \label{fig2}
\end{figure*}

aHydro versus RTA Boltzmann comparison is shown in Fig.~\ref{fig1}, with the shear-viscosity-to-entropy ratio being $\bar{\eta} = \frac{5}{4\pi}$ and initial parameters $\Lambda_0 = 1/\tau_i$ and $\xi_0 = 1$. Orange lines are the numerical results of aHydro. The gray lines represent the asymptotic formulae~\eqref{eq:asymptotic}, which approach the full solutions at late time $\hat\tau \gtrsim 100\,\tau_i$. The convergence of orange and gray lines at late time validates the applicability of the asymptotic formulae~\eqref{eq:asymptotic}, which provides intuition of how the solution approaches the modified Bjorken solution. Additionally, physics observables, i.e., temperature and pressure anisotropy, are presented in the lower two panels of ~Fig.~\ref{fig1} in comparison with the RTA Boltzmann results~(\ref{eq:RTA:temperature}, \ref{eq:RTA:pressure}) with identical initial condition. No visible difference is observed which validates that aHydro framework is a good approximation for the RTA Boltzmann equation taking the RS initial conditions. Such comparison is consistent with corresponding results in boost-invariant systems~\cite{Strickland:2017kux, Strickland:2018ayk}.

\section{Attractor of aHydro}\label{sec:attractor}
For systems that are boost-invariant and homogeneous in the longitudinal and transverse directions, attractor solutions of the aHydro equations have been found~\cite{Strickland:2017kux}. Naturally, one would expect the same conclusion for systems with more general rapidity structure. It is straight forward to see that such attractor solutions are applicable in the current system, by simply performing the mapping that
\begin{align}
\begin{split}
    &\Big\{\tau,\; \xi(\tau),\; \Lambda(\tau),\; T(\tau), \; \frac{P_\mathrm{L}(\tau)}{P_\mathrm{T}(\tau)}\Big\}
\\\to
    &\Big\{\hat{\tau},\; \xi(\hat{\tau}),\; \Lambda(\hat{\tau}),\;
    \frac{\hat{T}(\hat{\tau})}{\sqrt{\hat{g}^{\tau\tau}}},\;
    \frac{\hat{P}_\mathrm{L}(\hat{\tau})}{\hat{P}_\mathrm{T}(\hat{\tau})}\Big\}\,.
\end{split}
\label{mapping}
\end{align}
Therefore, we define
\begin{align}
    \hat{w} 
\equiv\,& 
    \hat{\tau} 
    \frac{\hat{T}(\hat{\tau})}{\sqrt{\hat{g}^{\tau\tau}}}
=
    \hat{\tau}\, \mathcal{R}^{\frac{1}{4}}(\xi(\hat{\tau})) \Lambda(\hat{\tau})\,,\label{eq:attractor:w}\\
\hat\varphi
\equiv\,& 
    \hat{\tau} \frac{\hat{\partial}_{\tau} \hat{w}(\hat{\tau})}{\hat{w}(\hat{\tau})}
=
    1 + \hat{\tau} \frac{\hat{\partial}_{\tau} \hat{T}(\hat{\tau})}{\hat{T}(\hat{\tau})}\,,
\end{align}
and it is straight forward to see that $\hat{\tau}\hat{\partial}_{\tau} = \hat{\varphi}\, \hat{w}\partial_{\hat{w}}$.
Taking the new variables, one finds that Eq.~\eqref{eq:aHydro_eq2} becomes a constrain equation,
\begin{align}
    \hat\varphi = 1+\frac{1+\xi}{2}\frac{\mathcal{R}'(\xi)}{\mathcal{R}(\xi)}\,,
\end{align}
and the pressure anisotropy~\eqref{eq:ahydro:pressure} can be simplified as $\frac{\hat{P}_\mathrm{L}}{\hat{P}_\mathrm{T}} = \frac{3-4\hat\varphi}{2\hat\varphi-1}$, whereas Eq.~\eqref{eq:aHydro_eq1} leads to a first-order, single-variable differential equation of $\xi(\hat w)$,
\begin{align}
\hat{w} \partial_{\hat{w}}\xi =\;& 
    \Big(2 -\xi(1+\xi)^{\frac{1}{2}}\mathcal R^{\frac{5}{4}}(\xi) \frac{\hat{w}}{5\bar\eta}\Big) \Big/
    \Big(\frac{1}{1+\xi}+\frac{\mathcal{R}'(\xi)}{2\mathcal{R}(\xi)}\Big)\,.
\end{align}
Finally, the $\tau$ and $\eta$ dependence can be obtained through $\hat{w}(\hat{\tau}(\tau, \eta))$, and we focus on the pressure anisotropy (${\hat{P}_\mathrm{L}}/{\hat{P}_\mathrm{T}}$) versus scaled time $\bar{w} \equiv \frac{\tau T(\tau,\eta)}{5\bar\eta}$ at different rapidity slices.

\begin{figure}[!hbpt]
    \centering
    \includegraphics[width=0.9\linewidth]
    {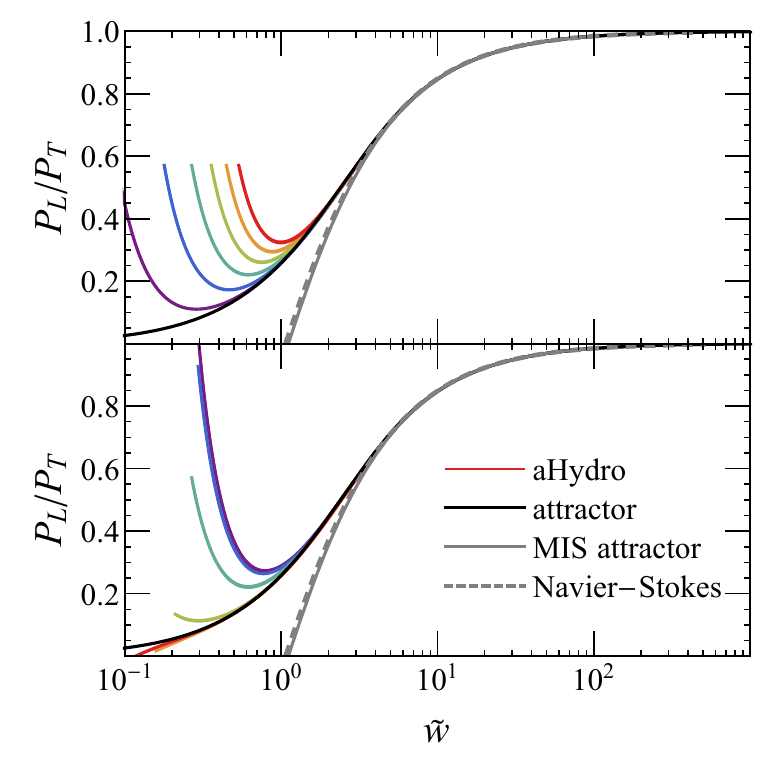}
    \caption{AHydro solution of $P_\mathrm{L}/P_\mathrm{T}$ (solid rainbow lines) in comparison with Attractors of aHydro equations (black solid), M\"uller--Isreal--Stewart hydro equation (gray solid), and Navier--Stokes solution (gray dashed). For rainbow curves, the upper panel is varying $\xi_0$ as Fig.~\ref{fig2} (left) and lower panel is varying $\Lambda_0$ as Fig.~\ref{fig2} (right).}
    \label{fig4}
\end{figure}
In Fig.~\ref{fig2} we show results from the aHydro equations~(\ref{eq:aHydro_eq1}, \ref{eq:aHydro_eq2}) and the attractor, and we introduce the rapidity dependence by letting $t_0=0.01\,\tau_i$. For different rapidity slices, we observe that varying different initial conditions of $\Lambda_0$ or $\xi_0$, the pressure anisotropy agrees with those from Boltzmann equation and both of them approach the attractor solution before global equilibrium. We also observe that at different $\eta$ the time of convergence can be different by a factor of $\sim 3$.  
Although this is understandable since large $\eta$ region is lower in temperature and requires longer time to be equilibrated, the difference in convergence time calls for a refined comparison with the attractor. This can be done by redefining the attractor variables into a manner that is invariant under Lorentz transformation, as previously introduced the same authors of the current paper~\cite{Chen:2024pez},
\begin{align}
    \tilde{w} \equiv \frac{1}{5\bar{\eta}}\frac{T}{\theta},\qquad
    \tilde \varphi \equiv 1+\frac{u^\mu \mathcal{D}_\mu T}{\theta\,T}\,,
    \label{eq:novel_attractor}
\end{align}
where $\theta = \mathcal{D}_\mu u^{\mu}$ represents the expansion rate of the system. It would return to the original definition for Bjorken flow when $t_0=0$, and one can further find $\tilde{w} = \frac{\hat{w}}{5\bar{\eta}}$ as defined in Eq.~\eqref{eq:attractor:w}. Fig.~\ref{fig4} shows the rapidity independent attractor behavior under different initial conditions of $\xi_0$ and $\Lambda_0$, in comparison with M\"uller--Isreal--Stewart attractor and Navier--Stokes solutions (see e.g. Ref.~\cite{Chen:2024pez} for details). It is evident that the aHydro solutions approach to their attractor before the convergence of aHydro and MIS attractors, which indicate the emergence of universal behavior of the aHydro equations before it approaches isotropization.

\section{Summary}\label{sec:summary}
In this work, we obtain the anisotropic hydrodymanics (aHydro) equations for systems that break the boost-invariance in the longitudinal direction. Time evolution of the temperature and pressure anisotropy are computed and good consistence is observed in comparison with the corresponding results from Boltzmann equation under relaxation time approximation.
We further obtain the corresponding attractor for aHydro equations and observed a rapidity dependent convergence time, if one makes the comparison with respect to the conventional attractor observables. We further show that after a redefinition of attractor observables, the convergence to the attractor is independent of rapidity, which happen earlier than their convengence with the M\"uller--Isreal--Stewart attractor and Navier--Stokes solutions. Such observations show the emergence of common behaviors in the distribution function when it approachs isotropization.

\vspace{5mm}
\emph{Acknowledgement.} The authors are grateful to Micheal Strickland for useful discussions. This work is supported by Tsinghua University under grant Nos. 53330500923 and 100005024.

\end{document}